\documentclass[aps,twocolumn,description,aps]{revtex4}
\usepackage{graphicx}
\usepackage{epstopdf}
\usepackage{amsmath}
\usepackage{amssymb}
\usepackage{multirow}
\usepackage{CJK}
\usepackage{color}
\usepackage[german,english]{babel}

\usepackage{color}

\begin{document}

\title{Nematic antiferromagnetic states in bulk FeSe}

\author{Kai Liu$^{1}$}
\author{Zhong-Yi Lu$^{1}$}\email{zlu@ruc.edu.cn}
\author{Tao Xiang$^{2,3}$}\email{txiang@iphy.ac.cn}

\affiliation{$^{1}$Department of Physics and Beijing Key Laboratory of Opto-electronic Functional Materials $\&$ Micro-nano Devices, Renmin University of China, Beijing 100872, China}
\affiliation{$^{2}$Institute of Physics, Chinese Academy of Sciences, Beijing 100190, China}
\affiliation{$^{3}$Collaborative Innovation Center of Quantum Matter, Beijing, China}

\date{\today}

\begin{abstract}
We revisit bulk FeSe through the systematic first-principles electronic structure calculations. We find that there are a series of staggered $n$-mer antiferromagnetic (AFM) states with corresponding energies below that of the collinear AFM state which is the ground state for the parent compounds of most iron-based superconductors. Here the staggered $n$-mer ($n$ any integer $>1$) means that a set of $n$ adjacent spins parallel on a line along $b$-axis with spins in antiparallel between $n$-mers and along $a$-axis. Among them, the lowest energy states are quasi-degenerate staggered dimer and staggered trimer AFM states as well as their any staggered combinations. Thus, to have the largest entropy to minimize the free energy at low temperature, the most favorable state is such a quasi-one-dimensional antiferromagnet in which along $b$-axis a variety of $n$-mers, mostly dimers and trimers, are randomly antiparallel aligned while along $a$-axis spins are antiparallel aligned, i.e. actually a nematic paramagnet. This finding accounts well for the absence of long-range magnetic order in bulk FeSe and meanwhile indicates the dominant stripe spin fluctuation and the nematicity as spin-driven.
\end{abstract}


\maketitle

\underline{Introduction:}
Recently there is a surge of interest on the investigation of physical properties of bulk FeSe, which is an 8 K iron-based superconductor at ambient pressure \cite{11}. A peculiar and yet puzzling phenomenon revealed by the experiments is that this material undergoes a nematic structural phase transition around 90 K \cite{mcqueen09prl} but without developing any kind of long-range magnetic orders \cite{mcqueen09prl}. The absence of magnetic order in FeSe provides a unique opportunity to explore the origin of the nematic phase and the role of antiferromagnetic (AFM) fluctuations in the superconducting pairing. This puts strong constrain on the theory of iron-based superconductivity, if the pairing mechanism in FeSe is not different from all other iron-based superconductors.

The nematic and AFM orders behave like a pair of twin brothers. They were observed in nearly all parent compounds of iron pnictides \cite{DaiPC08,HuangQ08}, FeTe \cite{BaoW09} or other iron-chalcogenides \cite{BaoW11}, and considered to be the two key factors affecting the superconducting pairing, since the iron-based superconductivity emerges after these orders are partially or completely suppressed by doping or pressure. As the nematicity already exists above the structural transition temperature $T_s$ \cite{Nakayama14}, it is believed that the nematic order is driven predominantly by electronic interactions rather than lattice interactions. Among the electronic mechanism, two scenarios have been proposed to explain the origin of the nematicity. One is to attribute the nematic order as a quasi-one-dimensional ferromagnetic orbital order of Fe 3$d_{xz}$ and 3$d_{yz}$ states and take the orbital fluctuations as the driving force for the nematic phase \cite{KrugerF09,Lee09,Onari12}. The other is to associate the structural transition with the formation of spin nematic order driven by the spin fluctuations, based on the analysis of the $J_1$-$J_2$ spin Heisenberg model \cite{FangC08,Xu08}. Since the orbital and spin fluctuations favor respectively the sign-preserving and sign-changing pairing \cite{Fernandes14}, to clarify the origin of nematic state is critical to the understanding of microscopic mechanism of iron-based superconductivity.

The nematic transition of bulk FeSe is clearly not directly related to long-range magnetic ordering. However, strong AFM spin fluctuations were observed  by neutron scattering \cite{RahnPRB15, zhaojunarxiv15} and nuclear magnetic resonance (NMR) \cite{Imai09} experiments, not just in the low temperature nematic phase, but also in the tetragonal phase above $T_s$ in this material \cite{zhaojunarxiv15}. Moreover, as revealed by inelastic neutron scattering measurements, the stripe spin fluctuation is strongly enhanced just below $T_s$ \cite{zhaojunarxiv15}. From muon-spin rotation ($\mu$SR) measurement, it was also found that a static AFM order emerges under an external pressure about 1 GPa \cite{Bendele10}. In addition, a sharp spin resonance was observed in the superconducting state \cite{zhaojunarxiv15}, which implies that the superconducting pairing in FeSe possesses a spin fluctuation-mediated sign-changing pairing symmetry. These experimental findings suggest that there are strong AFM interactions in FeSe, and the absence of magnetic orders is probably due to the AFM fluctuations and their interplay with other physical effects.

In order to understand why the magnetic order is absent in FeSe, a number of theories based on the calculations of extended Heisenberg or multi-orbital Hubbard models have been proposed. Using a two-orbital Hubbard model, Wang and Nevidomskyy \cite{WangZT15} found that there exists a parameter region with  moderately large electron doping where the orbital nematic order can survive without long-range magnetism. On the other hand, Wang, Lee and Kivelson argued that FeSe is a nematic quantum paramagnet sandwiched between a quantum N\'eel order phase and a striped AFM phase, and the nematic phase in FeSe is driven primarily by magnetism \cite{WangFarxiv15}. Furthermore, from the extended spin Heisenberg model with the bilinear-biquadratic interactions up to the third nearest neighbors, it was also found that the spin fluctuation can suppress magnetic but not nematic order formed either by ferro-orbital order \cite{glasbrenner15} or antiferroquadrupolar Ising-nematic order \cite{YuRarxiv15}. The Ising-nematic state can also emerge from the strong competition between the spin-density-wave and charge-current density-wave interaction channels induced by the small Fermi pockets in FeSe \cite{Chubukov15}.
However, all of the theoretical studies on the paramagnetic nematicity in FeSe are based on effective models, direct evidence from the first-principles electronic structure calculations is still lacking \cite{glasbrenner15}.

\begin{figure*}[!ht]
\includegraphics[width=15.5cm]{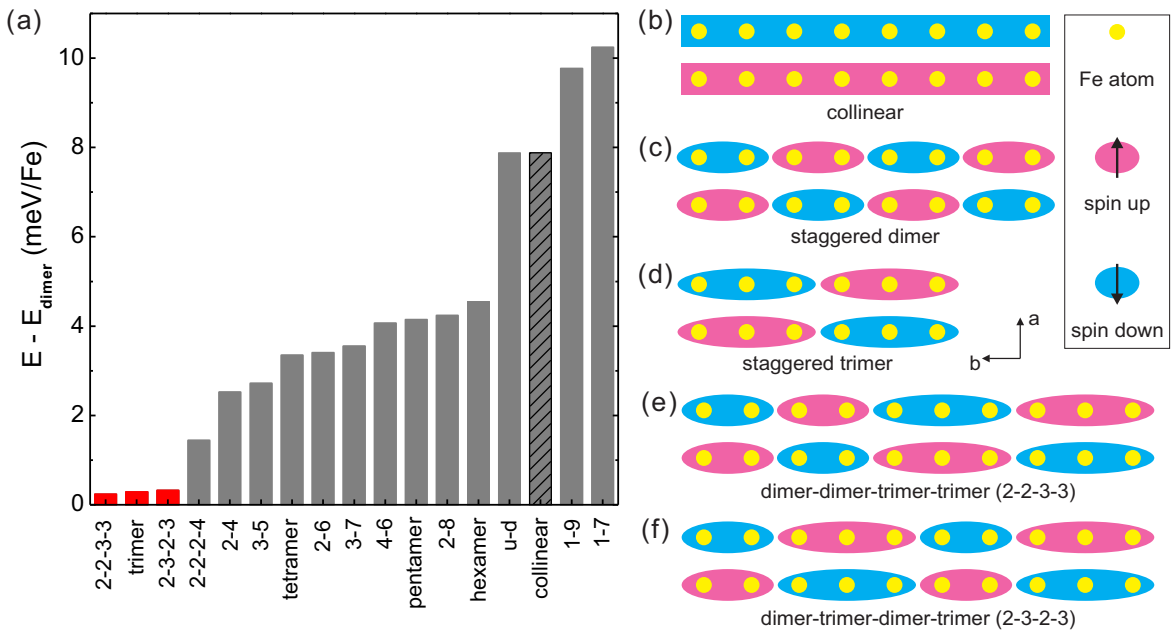}
\caption{(Color online)
  (a) Relative energies of various antiferromagnetic (AFM) states with respect to the staggered dimer AFM state for bulk FeSe. Spin configurations: (b) the collinear AFM (collinear), (c) the staggered dimer (dimer), (d) the staggered trimer (trimer), (e) the dimer-dimer-trimer-trimer (2-2-3-3), and (f) the dimer-trimer-dimer-trimer (2-3-2-3) states. Here the Fe atoms are denoted by yellow dots. The magenta and cyan areas represent the up and down spin, respectively. The directions of the $a$ and $b$ axes are shown in panel (d). The ``tetramer", ``pentamer", and ``hexamer" represent the staggered tetramer, staggered pentamer, and staggered hexamer AFM states in panel (a), respectively. The other spin configurations whose relative energies are given in panel (a) are shown in Fig. S1 of the Supplementary Materials.
}
\label{fig1}
\end{figure*}

\underline{Results and analysis:}
In this paper, we perform a systematical study on the electronic and magnetic structures of bulk FeSe through the first-principles density functional theory calculations. Details of the computational method and parameters used are given in Sec. A of the Supplemental Materials.
The ground state of bulk FeSe is found to be a pair-checkerboard AFM state, or the staggered dimer state shown in Fig. \ref{fig1}(c), which agrees with the result previously published by Cao \textit{et al.} \cite{cao15}, but disagrees with the experimental observation.

In order to see how the staggered dimer AFM order is melted by quantum or thermal fluctuations, we enlarged the unit cell and calculated the electronic structures of different AFM states, including the staggered trimer state as well as other AFM states as shown in Fig. \ref{fig1}(b-f) and Fig. S1 in Sec. B of the Supplemental Materials.
Here we conjectured that there should be a number of magnetic states energetically between the staggered dimer AFM state and the collinear AFM state. Very likely, such states are the other staggered $n$-mer ($n>$ 1) AFM states. Here the $n$-mer means that a set of $n$ adjacent spins on a line are parallel aligned.
As a result, Fig. \ref{fig1}(a) shows the relative energies of these states with respect to the staggered dimer state. Noticeably, the energy difference between the staggered dimer and trimer AFM states is very small, less than 0.3 meV/Fe (equivalent to about 3 K), consistent with the result reported in Ref. [\onlinecite{glasbrenner15}]. This energy difference is negligibly small in comparison with the magnetic interaction, the spin-orbital coupling and other energy scales of FeSe. In particular, the quantum spin fluctuations of the staggered dimer and trimer states may further reduce their energy difference. This suggests that the staggered dimer and trimer states can be approximately regarded as degenerate if the measurement temperature is well above their energy difference.

Nevertheless, the energy of the staggered $n$-mer AFM state ($n>1$) increases with $n$ [Fig. \ref{fig1}(a)]. The energy difference between the staggered dimer and tetramer states is about 3.4 meV/Fe, which is one order of magnitude higher than the energy difference between the staggered dimer and trimer states. For the other staggered $n$-mer states with $n>4$, the difference is even larger. The energy of the collinear AFM state is about 7.9 meV/Fe higher than the staggered dimer state. Clearly, there is a big energy gap between the staggered dimer or trimer states and other staggered $n$-mer states.

Besides the equal-length staggered $n$-mer states, we also studied the staggered ($n$-mer)-($l$-mer) ($m\not=l$) AFM states (Fig. S1).
The results demonstrate that once these states contain the monomer, the corresponding energies are higher than that of the collinear AFM state. Otherwise, the energy is below the collinear AFM state but higher than the staggered trimer state [Fig. \ref{fig1}(a)].

Considering the nearly degeneracy between the staggered dimer and trimer states, we studied more complex combinations among the staggered dimers and trimers. In particular, within a $\sqrt{2} \times 5\sqrt{2}$ supercell, the staggered dimer-dimer-trimer-trimer (2-2-3-3) [Fig. \ref{fig1}(e)], dimer-trimer-dimer-trimer (2-3-2-3) [Fig. \ref{fig1}(f)],  dimer-dimer-dimer-tetramer (2-2-2-4)[Fig. S1(j)], and 2-2-3-3up\_2-3-2-3down (u\_d) [Fig. S1(k)] states were examined, respectively. As shown in Fig. \ref{fig1}, the energies of the 2-2-3-3 and 2-3-2-3 AFM states are almost degenerate with those of the staggered dimer and trimer AFM states. The energy of the (2-2-2-4) AFM state lies almost at the middle of the energies of the staggered dimer and tetramer states. On the other hand, the energy of the u\_d AFM state is close to that of the collinear AFM state (Fig. \ref{fig1}). We also studied the 2-2-2-2-2-2up\_3-3-3-3down AFM state with a $\sqrt{2} \times 6\sqrt{2}$ supercell, and found that its energy is much higher than that of the collinear AFM state. Moreover, we calculated the system with dimers antiparallel aligned along $b$-axis and spins parallel aligned along $a$-axis, then found that it is energetically  about 52 meV/Fe higher than the collinear AFM state. Thus the low energy states of bulk FeSe are always antiferromagnetically coupled along $a$-axis.

\begin{figure}[!t]
\includegraphics[angle=0,scale=0.25]{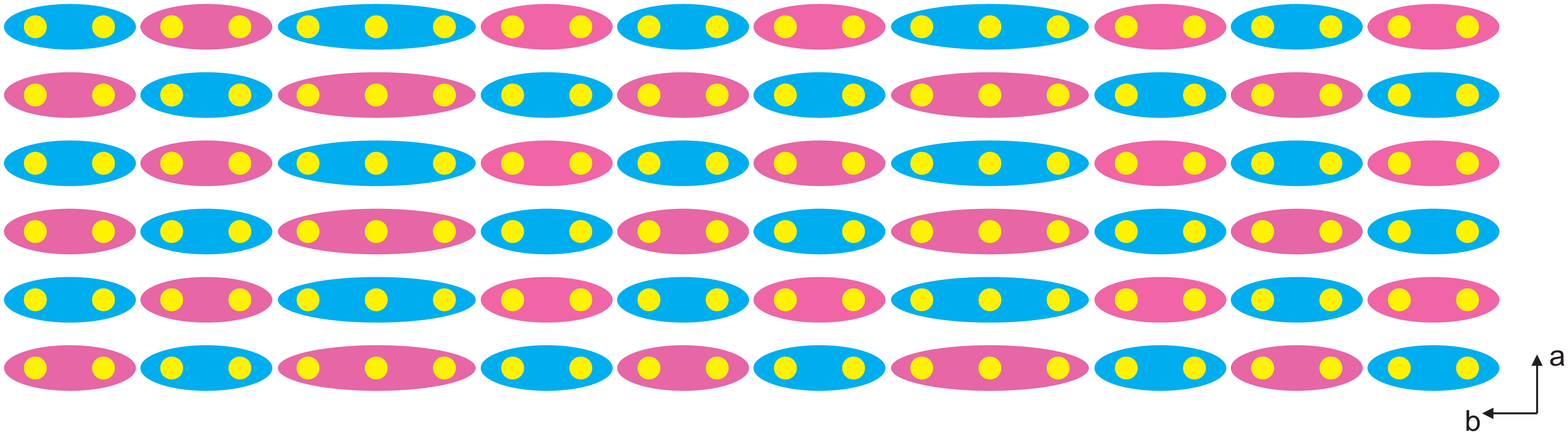}
\caption{(Color online)
  A typical spin configuration  in the lowest energy subspace of bulk FeSe. The spin correlation along the $a$-axis (vertical) is antiferromagnetically ordered, but along the $b$-axis (horizontal) a random staggered combination of spin dimers and trimers (2-2-3-2-2-2-3-2-2-2 here).
}
\label{fig2}
\end{figure}

The above results indicate that the staggered dimer and trimer states as well as their random staggered combinations along $b$-axis with spins AFM correlated along $a$-axis are nearly degenerate in a narrow window of energy (about 0.3 meV/Fe). They form a subset of states whose energies are well separated from other AFM states. Figure \ref{fig2} schematically shows a typical spin configuration of such an AFM state in this quasi-degenerate subspace. If the measurement temperature is higher than their energy differences (about 3 K), the probability to fall into any one of these states is nearly the same. In other words, these states will be strongly mixed by thermal and quantum fluctuations to gain the entropy to minimize the free energy. Due to the randomness in the distribution of dimers and trimers along $b$-axis, there is no any long-range magnetic order along this direction. Along $a$-axis, the spins are antiferromagnetically ordered. Thus the low energy states of FeSe are spin-nematic. But unlike usual quasi-one-dimensional AFM states, this order does not lead to a long-range magnetic order and any Bragg peak in the elastic neutron scattering spectra of FeSe, again due to the random distribution of dimers and trimers along $b$-axis.
This agrees with the experimental observations, and gives a natural explanation for the absence of long-range magnetic order in bulk FeSe at ambient pressure\cite{mcqueen09prl}.

In FeSe, the spins couple strongly with the orbital degrees of freedom. Thus the spin nematicity of the low energy states of FeSe will naturally lead to a nematic structural phase transition, in accordance with the experimental observations for the tetragonal-to-orthorhombic structural transition in bulk FeSe \cite{mcqueen09prl}. Similar to the collinear AFM state, we found that there is a small structural distortion in all the states of staggered dimer, trimer, or their arbitrary combinations. The lattice constant is slightly expanded along the AFM direction ($a$-axis) but slightly contracted along the $b$-axis. The ratio of the distortion, measured by the relative difference of the lattice constants along the $a$- and $b$-axes, $(a-b)/(a+b)$, is about $0.5\%$, which agrees quantitatively with the experimental results \cite{Margodonna08,Khasanov10}. This suggests that the nematicity observed in bulk FeSe below 90 K is mainly driven by magnetic interactions, consistent with the recent neutron scattering measurement \cite{zhaojunarxiv15}.

The quasi-degeneracy of the staggered dimer, trimer, and their arbitrary combinations along the $b$-axis suggests that there is a strong oscillation or mixing between two adjacent dimer and trimer and the coupling along this direction is very weak. This oscillation will be further enhanced by the quantum fluctuation of spins. However, as all the low energy states are AFM correlated along the $a$-axis, the AFM interaction along the $a$-axis must be very strong. Thus FeSe behaves more likely a quasi-one-dimensional spin ladder system with the width of the ladder equal to 2 or 3. Within each 2- or 3-leg ladder, the intra-ladder coupling is effectively ferromagnetic, but the inter-ladder coupling is effectively AFM. This suggests that the spin fluctuation in FeSe is highly one-dimensional, consistent with the picture suggested by Wang-Lee-Kivelson \cite{WangFarxiv15}. Moreover, thermal and quantum fluctuations can easily destruct the magnetic order due to this low dimensionality, leading to a paramagnetic phase as observed by experiments. Furthermore, the quasi-degeneracy of low energy states would imply that the electronic structure of FeSe is sensitive to the change of external pressure or crystal structure. Indeed, we found that the ground state of bulk FeSe is changed from the staggered dimer AFM state to the collinear AFM state by applying a pressure of 4 GPa.
This would explain why the superconducting transition temperature $T_c$ of FeSe is so susceptible to external pressures or chemical dopings: $T_c$ rises from 8 K at ambient pressure \cite{11} to 37 K under a pressure of 6-9 GPa \cite{Felser09,Margadonna09} or to above 40 K by the intercalation of alkali metal or other chemical elements between FeSe layers \cite{Guo10,Ying12,Burrard13,Lu14}.

Thus the quasi-degenerate AFM states formed by the staggered dimer, trimer, and their combinations behave like a nematic paramagnet on thermal average. Along the $a$-axis, these states are AFM ordered. But this ordering does not lead to any $\delta$-function like Bragg peaks in the elastic neutron scattering spectra, since the dimers and trimers along the $b$-axis are randomly distributed. Instead, it manifests as a broad peak on the axis of $k_a=\pi$ (the lattice constant is set to 1), with the center of the peak locating between the two characteristic wave vectors of the staggered trimer and dimer states along the $b$-axis, namely $k_b = \pi/3$ and $\pi/2$, respectively. As the energies of the staggered $n$-mer states ($n>3$) are close to that of the staggered dimer state, quantum fluctuations may mix some of these $n$-mer states with the dimer and trimer states. In such a case, the peak will be further broadened and the peak position will move towards the point $k_b =0$. This prediction for FeSe can be tested by the {\it elastic} neutron scattering measurement. One can also use a spin-resolved scanning tunneling microscope to directly detect the AFM order of the staggered dimer or other quasi-degenerate low energy states along the $a$-axis.

The above discussion indicates that the AFM correlations are very strong in FeSe, like in all other parent compounds of iron-based superconductors. The reason why FeSe behaves like a nematic paramagnet is mainly due to the strong thermal and quantum fluctuations in the low energy states, rather than the orbital ordering or other physical effects. This would explain why the first-principles electronic structure calculations succeed in determining the collinear AFM ground states of LaFeAsO \cite{Dong08,MaFJPRB08}, CaFe$_2$As$_2$ \cite{MaFJ10} and other '1111' and '122' materials as well as the bicollinear AFM ground state of FeTe \cite{MaFJPRL09}, but fail in the case of FeSe in the previous studies.

\begin{figure}[!t]
\includegraphics[angle=0,scale=0.35]{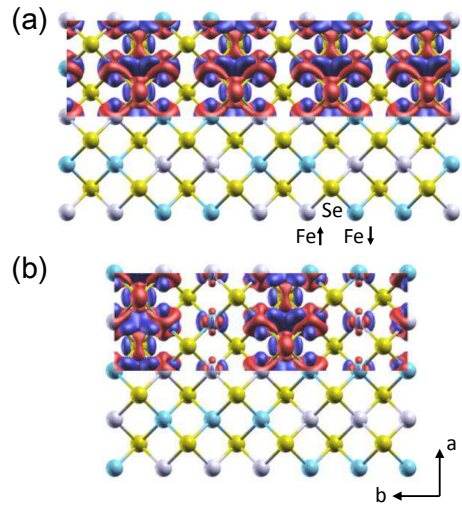}
\caption{(Color online)
  Difference of charge density between the staggered dimer and the collinear AFM states (a) and that between the staggered trimer and the collinear AFM states (b) for bulk FeSe. The lower and upper half part of each panel shows the crystal structure and the difference of charge density, respectively. Red and blue isosurfaces are the electron accumulation and depletion regions, respectively, at the isovalue of 0.005 e/\AA$^3$.}
\label{nicol-chargediff}
\end{figure}

Phenomenologically, the staggered dimer state can be understood by invoking the $J_1$-$J_2$-$J_3$ Heisenberg model of Fe moments \cite{cao15}. However, in order to understand why the energy difference between the staggered dimer and trimer states is so small, we found that this type of models, with or without the biquadratic interactions up to the third nearest neighbors, is not enough. To understand the physical origin for this, it is helpful to compare the charge distribution of the staggered dimer or trimer state with that of the collinear AFM state. Fig. \ref{nicol-chargediff} shows the difference of charge density between the staggered dimer or trimer and the collinear AFM states.
In comparison with the collinear AFM state, there is a strong charge redistribution in the staggered dimer or trimer state. In contrast, from a similar calculation we found that there is not such a strong charge redistribution in iron-pnictides. The redistribution happens mainly in the region between two Fe atoms with opposite magnetic moments along the $b$-axis in FeSe. In particular, the charge density is reduced between two Fe atoms with opposite spins along the $b$-axis. This reduces the direct overlap of Fe $3d$ orbitals and suppresses the ferromagnetic interaction between these two Fe spins. In the bonding area of the Fe and Se atoms along the diagonal direction, the charge density is also reduced. This weakens the bonding force between Fe and Se atoms and suppresses the AFM superexchange interaction between the two next-nearest-neighboring Fe spins bridged by Se orbitals. Thus there is a strong interplay between the charge dynamics and the magnetic interactions. This implies that the itinerant electrons should have a significant contribution to the magnetic correlation in FeSe.
This further explains why the simple $J_1$-$J_2$-$J_3$-type of Heisenberg model cannot account for the spin nematicity of bulk FeSe obtained from our calculations.

\underline{Summary:}
From the first-principles electronic structure calculations, we have shown that bulk FeSe is essentially a nematic paramagnet with quasi-one-dimensional AFM correlation along the $a$-axis in low temperatures. In particular, we found that there is a quasi-degenerate subset of AFM states with a very narrow energy window of just 0.3 meV/Fe, and their energies are significantly lower than other states. All these quasi-degenerate AFM states are nematic, since they are perfectly AFM ordered along the $a$-axis. Along the $b$-axis, these states are formed by the AFM arranged dimers, trimers, and their arbitrary combinations. Within each dimer or trimer, the spins are parallel aligned. On average, these states do not possess any long-range magnetic orders along the $b$-axis. Moreover, since the energy difference between the staggered dimer and staggered trimer states is so small, to change an adjacent dimer-trimer state to an adjacent trimer-dimer state along the $b$-axis costs a little energy. Thus this system is so strongly frustrated that thermal and quantum fluctuations can easily suppress the magnetic order along the $b$-axis, but leave the spin nematic order along the $a$-axis unchanged. The physical picture we obtained agrees well with the experimental observations. It reveals the origin of the nematicity and gives a natural account for the absence of long-range magnetic order in bulk FeSe.

\begin{acknowledgments}

We thank Dung-Hai Lee, Weiqiang Yu, and Jun Zhao for useful discussions. This work was supported by the National Natural Science Foundation of China (Grants No. 11190024 and 91421304). KL was supported by the Fundamental Research Funds for the Central Universities, and the Research Funds of Renmin University of China (14XNLQ03).

\end{acknowledgments}

\end{document}


\title{Nematic antiferromagnetic states in bulk FeSe}

\author{Kai Liu$^{1}$}
\author{Zhong-Yi Lu$^{1}$}\email{zlu@ruc.edu.cn}
\author{Tao Xiang$^{2,3}$}\email{txiang@iphy.ac.cn}

\affiliation{$^{1}$Department of Physics and Beijing Key Laboratory of Opto-electronic Functional Materials $\&$ Micro-nano Devices, Renmin University of China, Beijing 100872, China}
\affiliation{$^{2}$Institute of Physics, Chinese Academy of Sciences, Beijing 100190, China}
\affiliation{$^{3}$Collaborative Innovation Center of Quantum Matter, Beijing, China}

\date{\today}

\maketitle

\section*{Supplementary Materials}
Here we provide details of the computational method and parameters used in our calculations, and some of the spin configurations whose electronic structures and energies were evaluated.

\subsection{Computational method}\label{SI:CompMeth}

Our fully spin-polarized first-principles electronic structure calculations were carried out using the projector augmented wave (PAW) method \cite{paw1,paw2} as implemented in the VASP package \cite{vasp1,vasp2,vasp3}. The generalized gradient approximation of Perdew-Burke-Ernzerhof \cite{pbe} for the exchange-correlation potential was adopted. Due to the layered structure of bulk FeSe and its charge neutral character of each layer \cite{Ricci13,Nakamura13,Ye13}, the vdW-optB86b functional \cite{optb86} for the van der Waals correction was used. The kinetic energy cutoff of the plane-wave basis was set to 350 eV. The lattice constants were fixed at the experimental values ($a=b=3.773$ \AA, $c=5.526$ \AA) of the tetragonal phase of $\beta$-FeSe \cite{mcqueen09prb}. All internal atomic positions were relaxed until the forces were smaller than 0.01 eV/\AA.

We first studied the nonmagnetic, the checkerboard antiferromagnetic (AFM) N\'eel, and the collinear AFM states with a $\sqrt{2}\times\sqrt{2}$ supercell. For this supercell, a 12$\times$12$\times$12 {\bf k}-point mesh for the Brillouin zone sampling and the Gaussian smearing technique with a width of 0.05 eV were used. Then we studied the various staggered AFM states, whose spin configurations are shown in Fig. 1 in the main text and Fig. \ref{SI-fig} in this Supplementary Material. The supercells of these magnetic states are $\sqrt{2}\times N\sqrt{2}$ with $N$=3, 4, 5, and 6 respectively. The {\bf k}-point mesh along ${\bf k_b}$ direction were accordingly adjusted to 4, 3, 3, and 2, respectively. With the above parameters, the energy difference between the collinear AFM states calculated with different supercell sizes was found to be less than 0.05 meV/Fe.

Since only the internal atomic positions, not the lattice constants, were relaxed, the calculated energy difference (-7.9 meV/Fe) between the staggered dimer state and the collinear AFM state is smaller than the value (-15 meV/Fe) reported in Ref. [\onlinecite{cao15}].

Finally, we fully optimized the lattice structures of the staggered dimer and trimer states to check a possible tetragonal-to-orthorhombic structural phase transition.

\makeatletter
\renewcommand{\thefigure}{S\@arabic\c@figure}
\makeatother
\begin{figure*}[th]
\includegraphics[angle=0,scale=0.4]{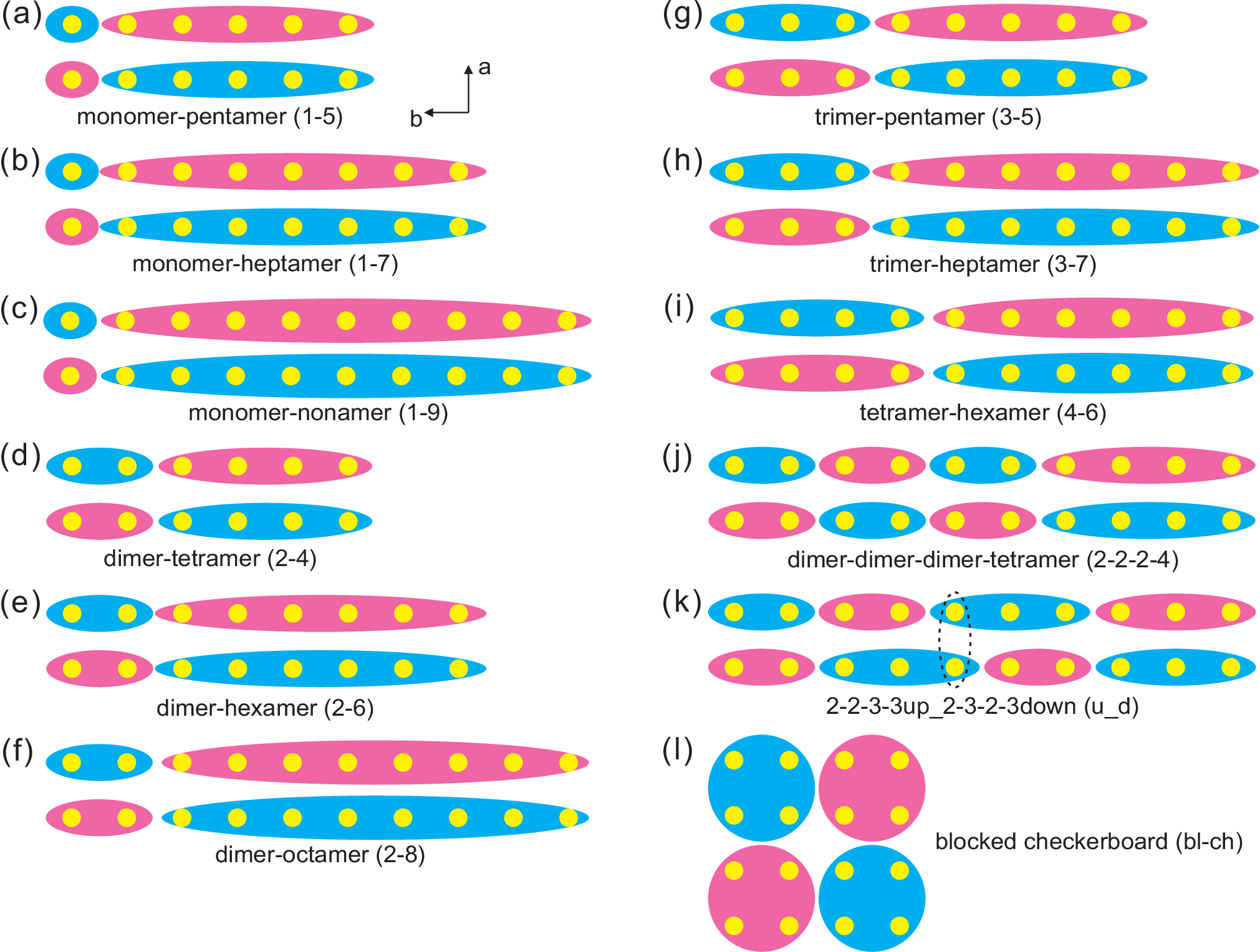}
\caption{(Color online)
  Spin configurations of bulk FeSe with the supercell length up to 10 Fe atoms along the $b$-axis (horizontal) direction whose relative energies with respect to the ground state energy of the staggered dimer state are shown in Fig. 1 in the main text. The Fe atoms are denoted by yellow dots. The magenta and cyan areas represent the up spin and down spin, respectively.
  The supercells used in the calculations are:
  $\sqrt{2}\times 3\sqrt{2}$ for the (1-5) and (2-4) states;
  $\sqrt{2}\times 4\sqrt{2}$ for the (1-7), (2-6), and (3-5) states;
  $\sqrt{2}\times 5\sqrt{2}$ for the rest of other states.
  A $2\sqrt{2}\times 2\sqrt{2}$ supercell was also used to investigate the blocked checkerboard AFM (bl-ch) state (l), namely an AFM state with dimers staggered along both $a$-axis and $b$-axis, whose calculated energy is about 80 meV/Fe higher than that of the staggered dimer AFM state.
}
\label{SI-fig}
\end{figure*}

\subsection{Magnetic configurations}
\label{SI-MagConf}

Fig. \ref{SI-fig} shows some spin configurations of bulk FeSe in the enlarged supercell whose electronic structures were calculated by the first-principles density functional theory. The relative energies of these states with respect to the staggered dimer state are shown in Fig. 1 in the main text.